\documentclass[showkeys,showpacs, aps,amssymb,keywords,floatfix,prd,amsmath,preprintnumbers]{revtex4}
\RequirePackage{float}
\usepackage{epstopdf}
\usepackage{hyperref}
\usepackage{capt-of}
\usepackage{graphicx}  % Include figure files
\usepackage{dcolumn}   % Align table columns on decimal point
\usepackage{bm}% bold math
\usepackage{booktabs}
\usepackage{nicefrac} 
\usepackage{mathtools}
\usepackage{resizegather}
\usepackage{xcolor}
\begin{document}
	
%\input epsf.tex
%%%%%%%%%%%%
%%%%%%%%%%%
\title{Configurational Entropy in Chaplygin Gas Models}

\author{Snehasish Bhattacharjee\footnote{Email: snehasish.bhattacharjee.666@gmail.com}
}

\affiliation{Department of Physics, Indian Institute of Technology, Hyderabad 502285, India}
\date{\today}

\begin{abstract}
The present work employs the Linder parametrization of a constant growth index \cite{linder/index} to investigate the evolution of growth rate of clustering and the dissipation of configurational entropy in some of the most widely studied Chaplygin gas models, such as the generalized Chaplygin gas and the modified Chaplygin gas. The model parameters of the Chaplygin gas models are found to play a vital role in the evolution of growth rate, dark energy density parameter, EoS parameter, and configurational entropy. Furthermore, the work communicate the rate of change of configurational entropy to attain a minimum which depend solely on the choice of model parameters and that there exists suitable parameter combinations giving rise to a viable dissipation of configurational entropy, and therefore certifying its time derivative to hit a minimum at a scale factor which complies with the current observational constraints on the redshift of transition from a dust to an accelerated Universe and thereby making Chaplygin gas models a viable candidate for dark energy.
\end{abstract}

\keywords{configurational entropy; chaplygin gas; observational constraints}

\pacs{04.50.Kd}

\maketitle

\section{Introduction}
Although the $\Lambda$CDM cosmological model elegantly explains the dynamics of the Universe, it predicts the existence of dark matter and dark energy in profuse quantities \cite{1} which regulates the gravitational collapse of matter and accelerated expansion of the Universe respectively. Cosmologists, therefore, are keen to understand the true nature of these entities to get a complete understanding of the cosmos. Unfortunately, both of these exotic forms of energy lacks observational confirmation and therefore challenge the principles of the standard cosmological model.  In addition to this, the so-called Hubble tension \cite{2} contributes to worsening the things and hints at either a completely new cosmological model or the existence of new fundamental physics.   \\
To alleviate these cosmological observations, numerous modified gravity theories and dynamical dark energy models have been proposed \cite{bull}. Chaplygin gas (CG) is one such promising candidate first proposed in \cite{cg24} with an EoS of the form 
\begin{equation}
p = -\frac{A}{\rho},
\end{equation} 
where $A>0$ is a constant. Although this particular model had some successes in cosmology, updated observations do not favor such an EoS. In due course, several modifications emerged with extra degrees of freedom to suffice the observations. One such model is the generalized Chaplygin gas (hereafter, GCG) \cite{cg25,cg26} with an EoS of the form
\begin{equation}
p = - \frac{A}{\rho^{\gamma}},
\end{equation} 
where $\gamma$ is a constant. Note that for $\gamma=-1$ and $A=1$ we end up with the cosmological constant. The idea behind the GCG model is that since it has two free parameters, they can always be tuned to make the model behave as a pressureless fluid at early times while as a cosmological constant at late times. In addition to the GCG model, yet another version of CG surfaced termed the modified Chaplygin gas (hereafter, MCG) with an EoS of the form \cite{cg27,cg28,cg29}
\begin{equation}\label{mcg}
p=B\rho-\frac{A}{\rho^{\gamma}},
\end{equation}
where $B$, $A$, and $\gamma$ are constants. Setting $B=0$ we obtain the GCG model while setting $A=0$, and $B=-1$ we get the cosmological constant. The MCG model although a single fluid model is interesting in the sense that it unifies both dark matter and dark energy and comply with many cosmological observations such as GRBs \cite{cg34} and gravitational lensing \cite{cg32,cg33}. \\
An interesting concept in cosmology is the recently proposed configurational entropy which states that the evolution of an almost perfect Gaussian matter distribution to the highly non-linear state could be due to the dissipation of configurational entropy \cite{19} and could also play a role in the current acceleration of the Universe \cite{20}. Configurational entropy is motivated from the fact that one must take into account the information entropy of the Universe in conjunction with other sources of entropy production and register the principle of the maximal entropy generation to obtain a better comprehension of the accelerated expansion of the Universe \cite{19}. Thus, in addition to the estimation of the entropies for several components of the Universe such as the CMB photons, neutrinos, baryonic content of the stars, ISM, IGM, stellar and supermassive black holes, one must also consider the change in entropy related to the evolution of the Universe from a linear state to the highly clumpy state \cite{19}. For a static Universe, the emptying of configurational entropy is expected as the presence of cosmic structures make the Universe unstable. Thus, on the assumption that the Universe represents a thermodynamic system as a whole, a direct consequence of the second law of thermodynamics requires the suppression of the growth rate of structures due to the cosmic acceleration owing to the non-existence of other entropy production processes to counterbalance the loss of configurational entropy \cite{20}. Another interesting feature of configurational entropy is that the dissipation of configurational entropy relies on the growth rate of density fluctuations and on the rate of the cosmic expansion. Thus, the rate of change of configurational entropy is a function of the underlying cosmological model \cite{19,20}. For a dust dominated Universe, the rate of change of configurational entropy is negative and because of which the Universe slows down and therefore structures begin to appear on largest of scales. As a consequence, configurational entropy dissipates at an enhanced rate under such conditions \cite{19,20}. The second law of thermodynamics warrants the fact that the rate of change of configurational entropy must always be less than the combined entropy production rates from all the sources \cite{19,20}. \\
Configurational entropy has been studied in the framework of teleparallel $f(T)$ gravity \cite{ft} and for constraining model parameters of Tsallis holographic dark energy \cite{tsallis}. In this work, we shall build upon this to constrain some Chaplygin gas models.\\
Throughout the work we use $\Omega_{m0}=0.315$, $\Omega_{\Lambda}=1-\Omega_{m0}$ and $h=0.674$ \cite{2} and work with natural units.

\section{Configurational entropy and growth rate }\label{sec2} 
Let us start with a region of volume $V$ in an isotropic and homogenous Universe. We shall dissect $V$ into a large number of smaller volume elements $dV$ accommodating energy density $\rho (\overrightarrow{x},t)$ where $\overrightarrow{x}$ and $t$ denote the comoving coordinates and time respectively.\\
The term configurational entropy ($\Psi (t)$) first defined in \cite{19} and based on information entropy \cite{shannon} reads
\begin{equation}
\Psi (t) = - \int \rho \hspace{0.05cm}\text{log} \hspace{0.05cm} \rho d V .
\end{equation}
Next, the equation of continuity for an expanding Universe reads
\begin{equation}\label{eq1}
\frac{\partial \rho}{\partial t}+ 3 \frac{\dot{a}}{a} \rho + \frac{1}{a} \bigtriangledown . (\rho \overrightarrow{\nu})  = 0,
\end{equation}
where $\overrightarrow{\nu}$  and $a$ denote the peculiar velocity of cosmological fluid and the scale factor in the volume element $dV$ respectively. \\
Next, multiplying Eq. \ref{eq1} with $(1+ \text{log} \rho)$, followed by an integration over $V$, yields \cite{19} 
\begin{equation}\label{1}
3 \frac{\dot{a}}{a} \Psi (a) + \frac{d \Psi(a)}{d a} \dot{a}  - \sigma (a)  = 0,
\end{equation}
where $\sigma (a) = \frac{1}{a} \int \rho (\overrightarrow{x},a) \bigtriangledown . \overrightarrow{\nu} d V$. \\
Now, the equation relating the density contrast ($\delta (\overrightarrow{x})$) at a given position $\overrightarrow{x}$, the growing mode of fluctuations ($D(a)$), and the divergence of peculiar velocity $\nu$ can be written as \cite{20} 
\begin{equation}\label{2}
a \dot{a} \frac{d D (a)}{d a} \delta (\overrightarrow{x}) =- \bigtriangledown.\nu(\overrightarrow{x}).
\end{equation}
Employing Eq. \ref{2} in Eq. \ref{1}, gives \cite{20}
\begin{equation}\label{3}
\frac{3}{a}( \Psi (a)-\Phi) + \frac{d \Psi (a)}{d a} +    \overline{\rho} f    \frac{ D^{2} (a)}{ a} \int\delta^{2}(\overrightarrow{x})dV  = 0,
\end{equation}
where $\Phi = \int \rho (\overrightarrow{x},a) dV  $ represents the mass contained within $V$, and 
\begin{equation}\label{growth}
f = \frac{d \text{ln}D}{d \text{ln}a}=  \Omega_{m}(a)^{\gamma}
\end{equation} 
is the dimensionless growth rate with $\gamma$ being the growth index and $\Omega_{m}(a)$ the matter density parameter. The growth index in GR is approxiamately 6/11 \cite{stein}. Since growth index for alternate theories of gravity and dark energy models usually contain the free parameters that characterizes the models and since observational data for $\gamma$ is readily available, it therefore acts as a powerful tool to constrain alternate theories of gravity and dark energy models \cite{louis}. The growth index $\gamma$ has been derived for $f(R)$ gravity \cite{huang}, $f(T)$ gravity \cite{spy}, and $f(R,T)$ gravity \cite{frt}. Additionally, several studies have investigated the consequences of constant and redshift dependent parametrizations of $\gamma$. In this work, we shall employ one of the most widely used parametrizations first proposed in \cite{linder/index} and takes the form
\begin{equation}\label{index}
\gamma = 0.55 + 0.05 (1+\omega (a=0.5)).
\end{equation}  
The evolution of configurational entropy is obtained from the numerical solution of Eq. \ref{3}. We set the initial condition $\Psi(a_{i}) = \Phi$ and set the time-independent quantities to unity.\\
The first derivative of configurational entropy reaches a minimum at a distinct scale factor $a_{DE}$ which depend on the selection of the model parameters and represents the epoch of dark energy domination \cite{20,tsallis}. It may be noted that the relative dominance of the dark energy model under consideration in controlling the growth rate of large scale structures dictates the location of $a_{DE}$ \cite{20}. Upon careful inspection, one may find that the third term Eq. \ref{3} contain a distinct combination of the scale factor, growing mode, and its temporal derivative and it is largely because of such conjunctions of these entities that the first derivative of configurational entropy attains a minimum and represent the eon of dark energy domination and therefore acts a robust and straightforward method to impose constraints on a cosmological model.\\
Ref \cite{ratra} imposed strict limits of the scale factor corresponding to the epoch of the current accelerated phase of the Universe. Therefore, the idea of the present work is to investigate whether a suitable choice of the parameters for the Chaplygin gas models allow the time derivative of configurational entropy to reach a minimum at the observationally consistent scale factor range.

\section{Cosmological parameters for the Chaplygin gas models}
The field equations in GR reads
\begin{equation}
T_{ij} = R_{ij}-\frac{1}{2}g_{ij}R,
\end{equation}
$R_{ij}$, $R$, $g_{ij}$ and $T_{ij}$ represent respectively the Ricci tensor, Ricci scalar, energy momentum tensor and the metric tensor in the usual 3+1 dimensions. For a flat FRW metric with ($'-','+','+','+'$) signature, we arrive at the following Friedmann equations 
\begin{equation}\label{friedmann1}
 \rho = 3\left[\frac{\dot{a}^{2}}{a^{2}} \right], 
\end{equation}
and
\begin{equation}\label{friedmann2}
- \rho = 2 \frac{\ddot{a}}{a}+\frac{\dot{a}^{2}}{a^{2}},
\end{equation}
where $p$ and $\rho$ denote the pressure and energy density of cosmic fluid respectively. \\
Substituting Eq. \ref{mcg} in Eq. \ref{friedmann1} and Eq. \ref{friedmann2}, the expression for the Hubble parameter and energy density read respectively as 
\begin{equation}
H(a)=H_{0}\left[\frac{\Omega_{m0}}{a^3}+(1-\Omega_{m0}) \left((1-A) \left(\frac{1}{a}\right)^{3 (B+1) (\gamma +1)}+A\right)^{\frac{1}{\gamma +1}}\right]^{0.5},
\end{equation}
and
\begin{equation}
\rho_{MCG}(a)=\Delta \left[A-(A-1) a^{-3 (B+1) (\gamma +1)}\right]^{\frac{1}{\gamma +1}},
\end{equation}
where $\Delta$ is an integration constant. Additionally, the EoS parameter $\omega_{MCG}$ is represented as
\begin{equation}
\omega_{MCG}(a)=B-A\left( \frac{ (B+1)}{A-(A-1) \left(\frac{1}{a}\right)^{3 (B+1) (\gamma +1)}}\right).
\end{equation}
\section{Results}
\begin{figure}[H]
\centering
  \includegraphics[width=7.5 cm]{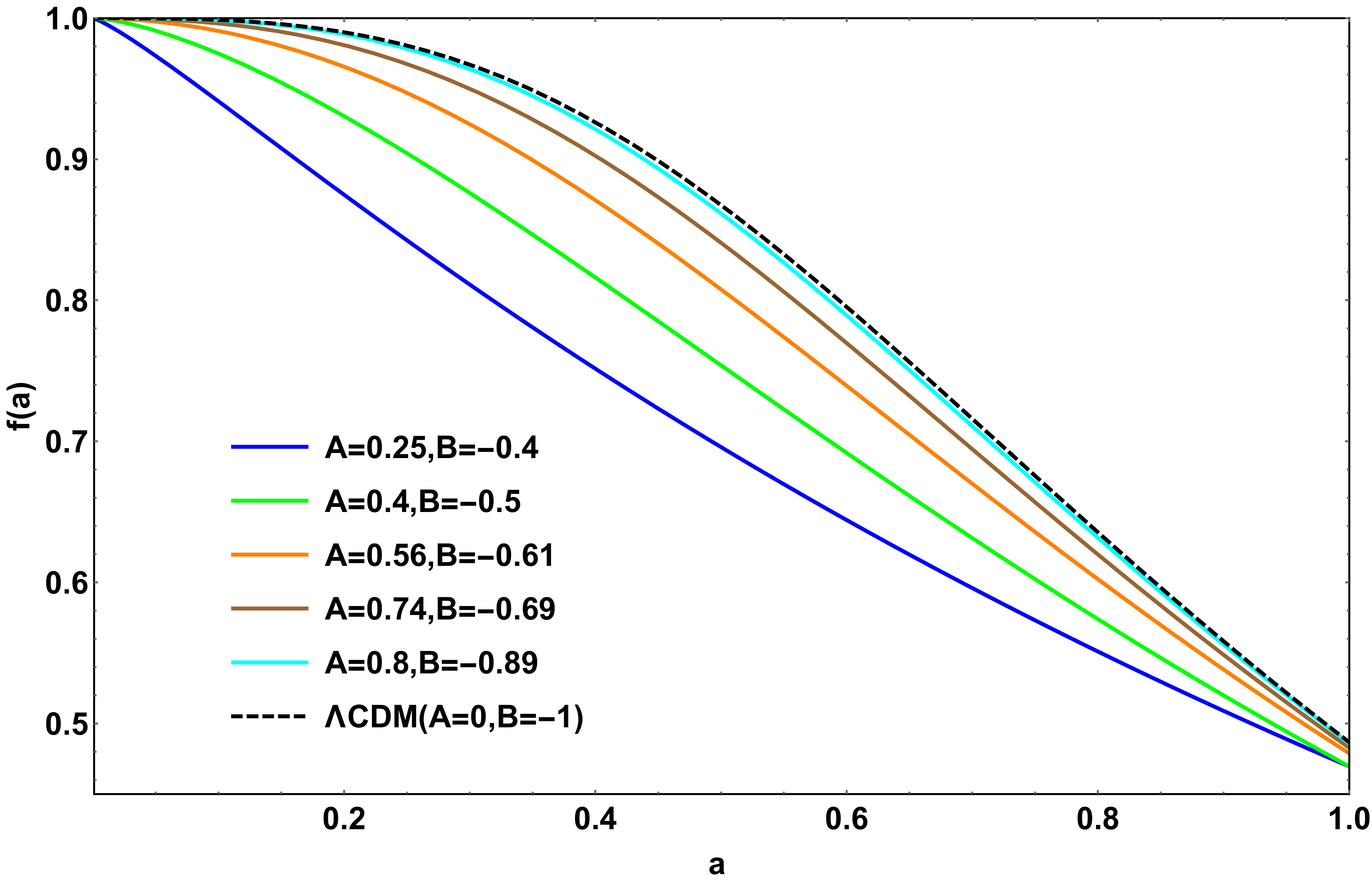}  
  \includegraphics[width=7.5 cm]{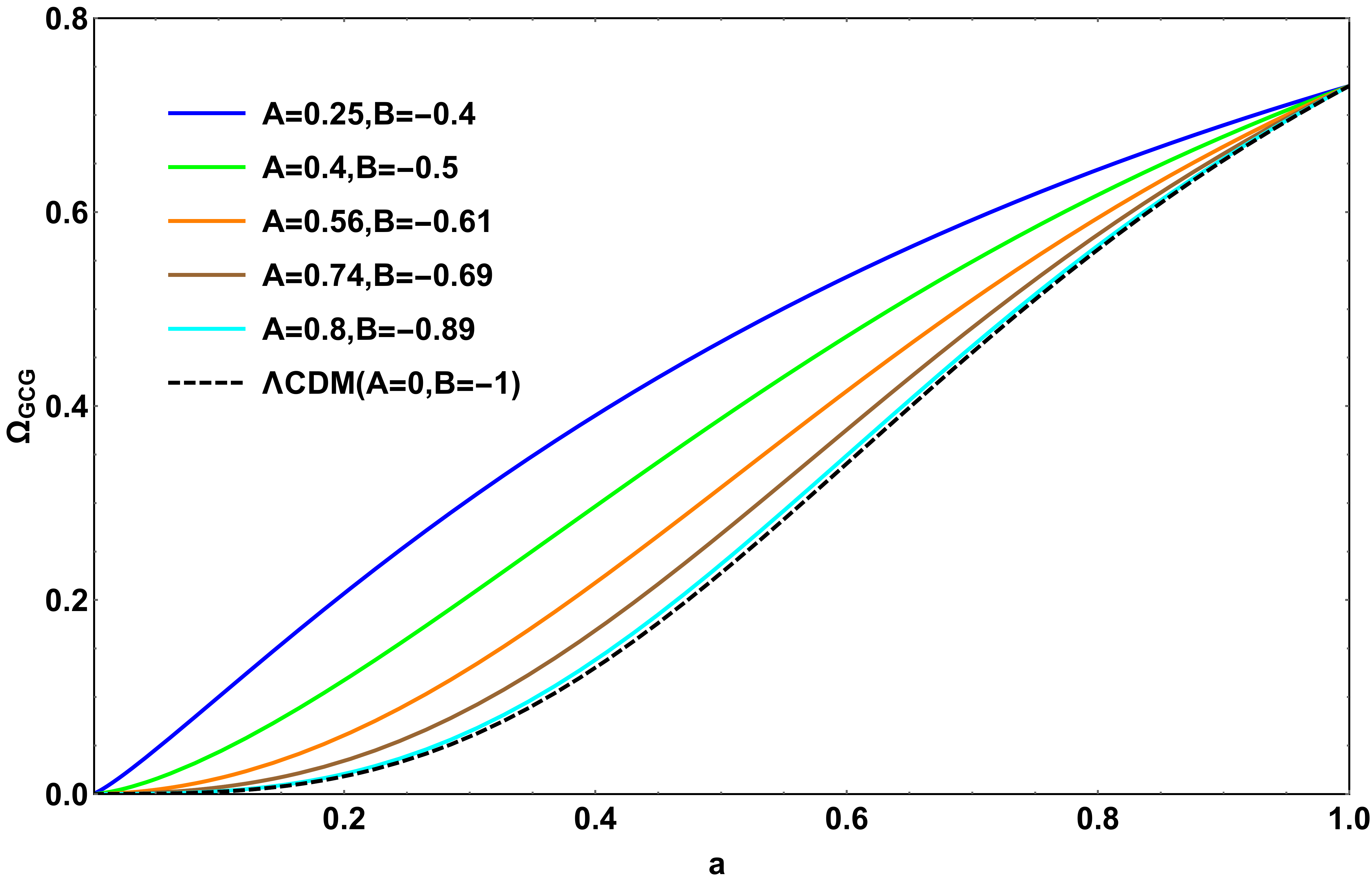}
  \includegraphics[width=7.5 cm]{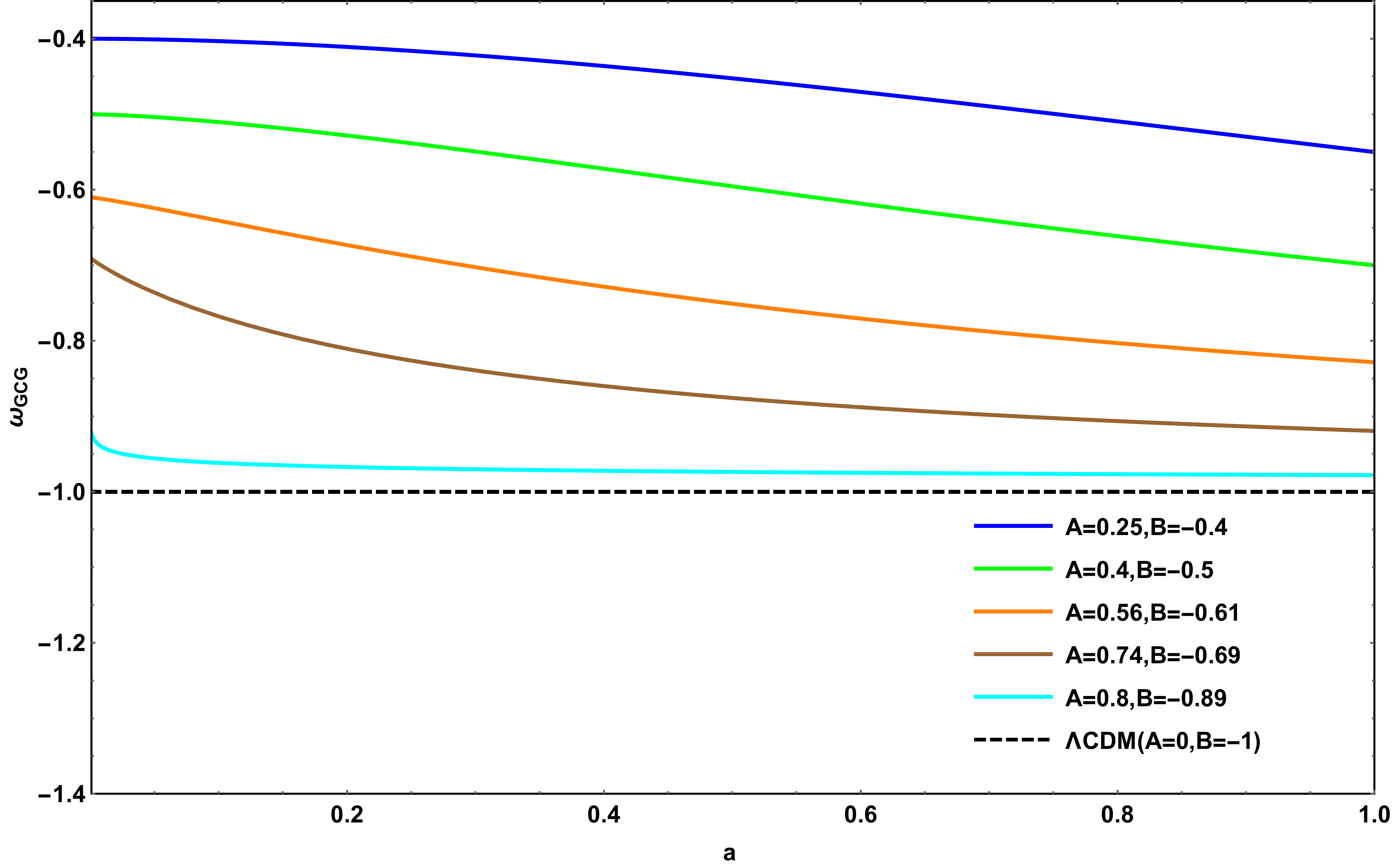}
  \includegraphics[width=7.5 cm]{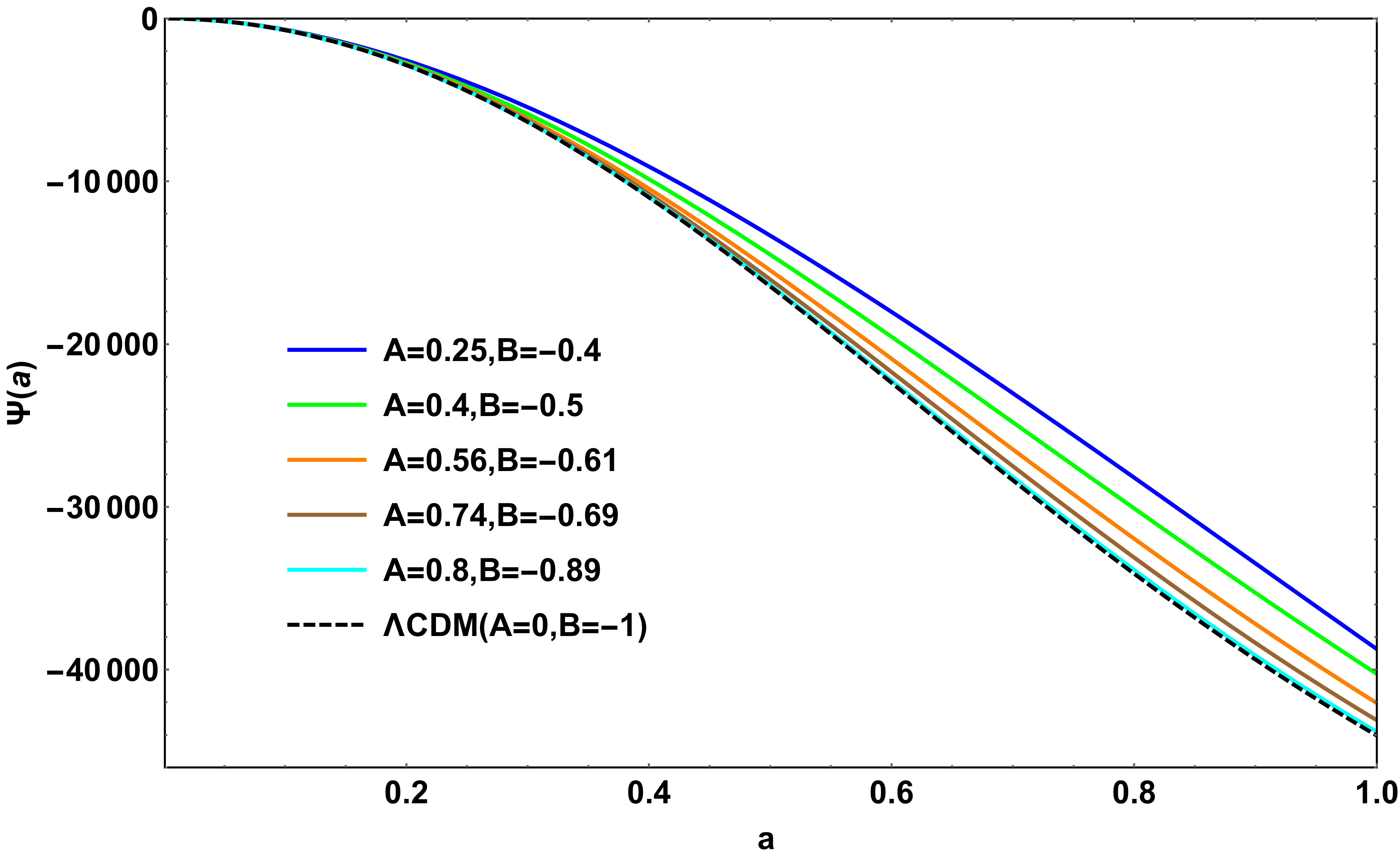}
  \includegraphics[width=7.5 cm]{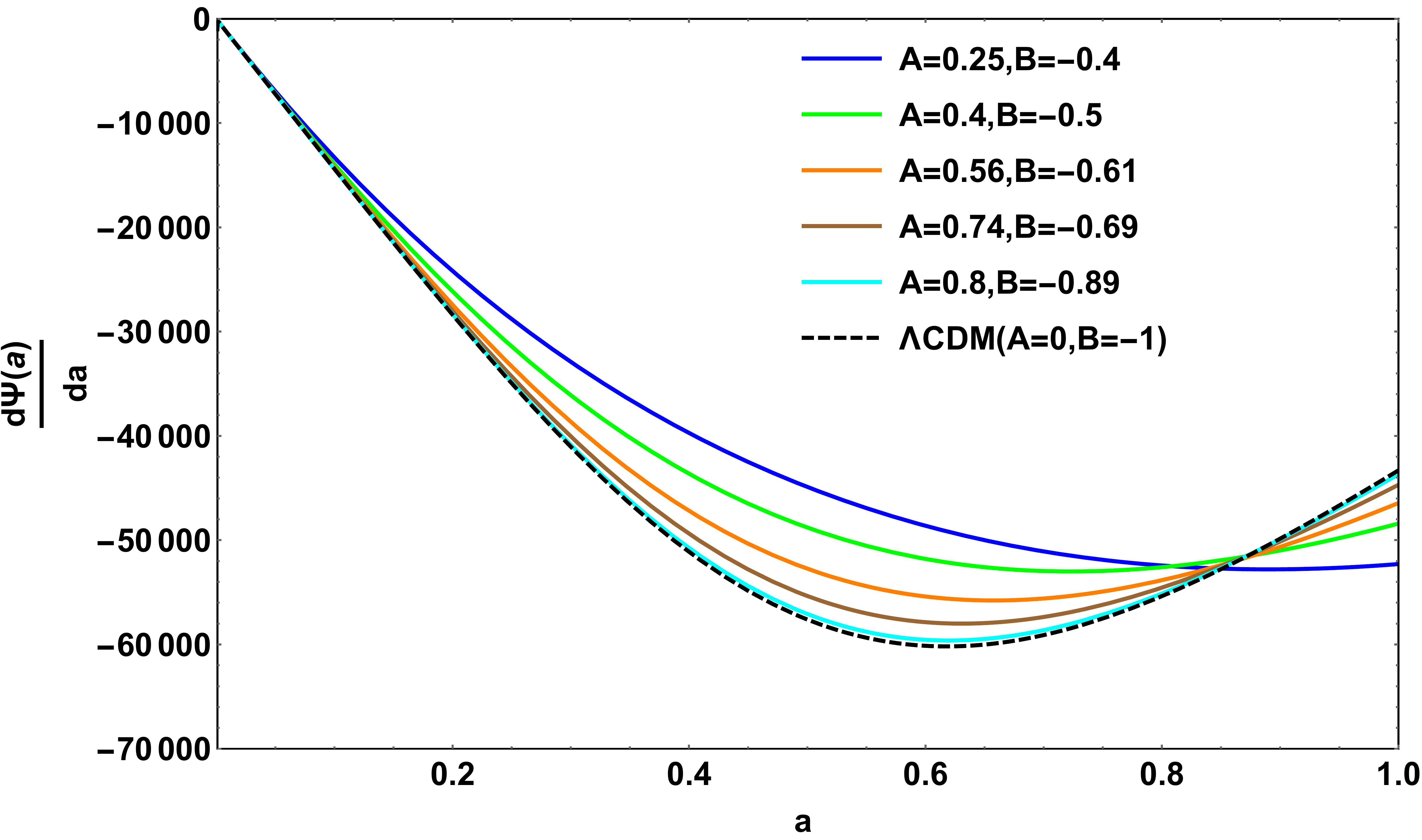}
\caption{Top left panel shows the evolution of growth rate $f(a)$, top right panel shows the evolution of dark energy density parameter $\Omega_{GCG}$, middle left panel shows the evolution of the EoS parameter $\omega_{GCG}$, middle right panel shows the evolution of the configurational entropy $\Psi(a)$, and the bottom left panel shows the rate of change of configurational entropy ($\frac{d\Psi(a)}{da}$) for the GCG model.}
\label{FIG1}
\end{figure}

\begin{figure}[H]
\centering
  \includegraphics[width=7.5 cm]{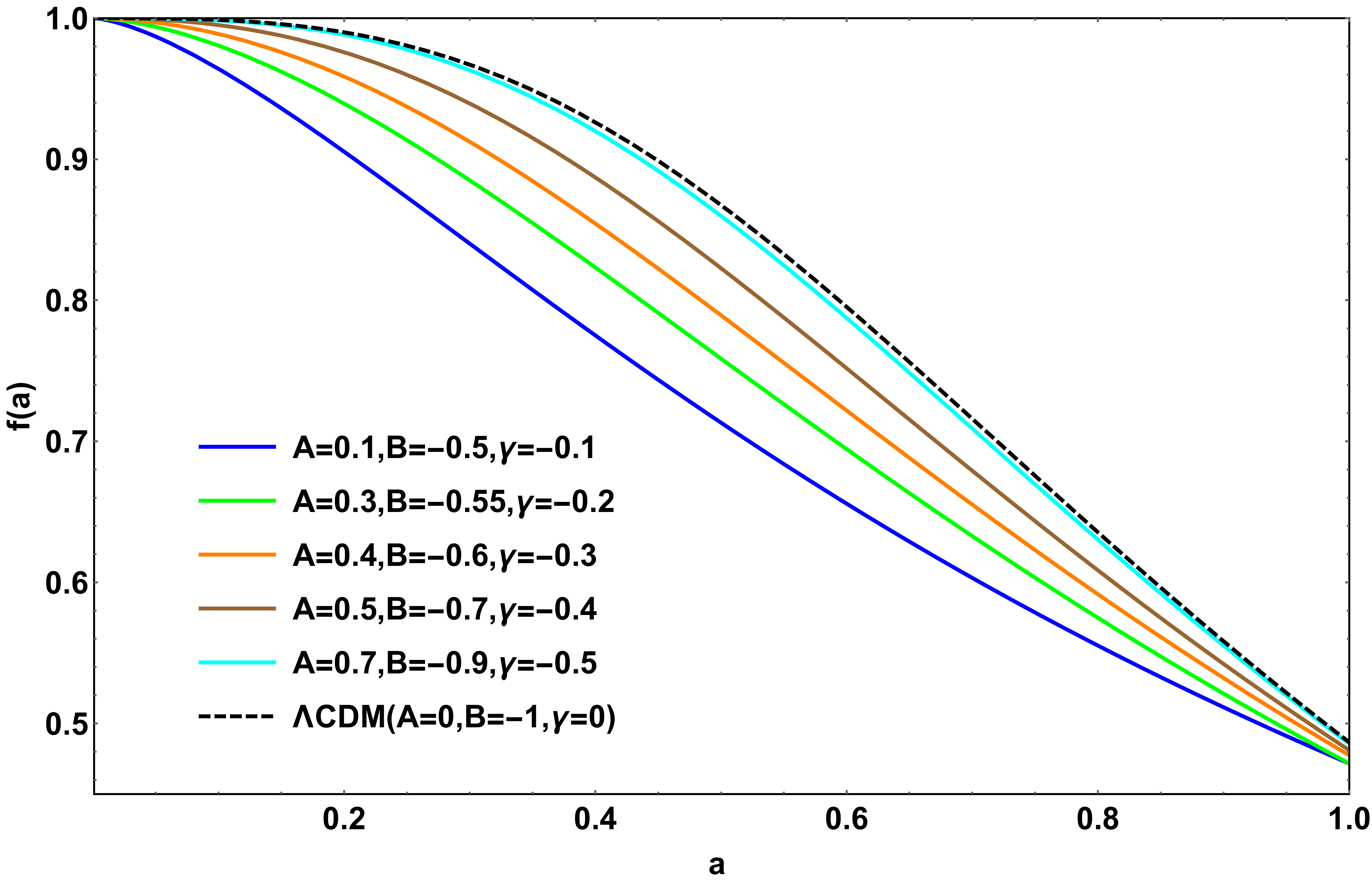}  
  \includegraphics[width=7.5 cm]{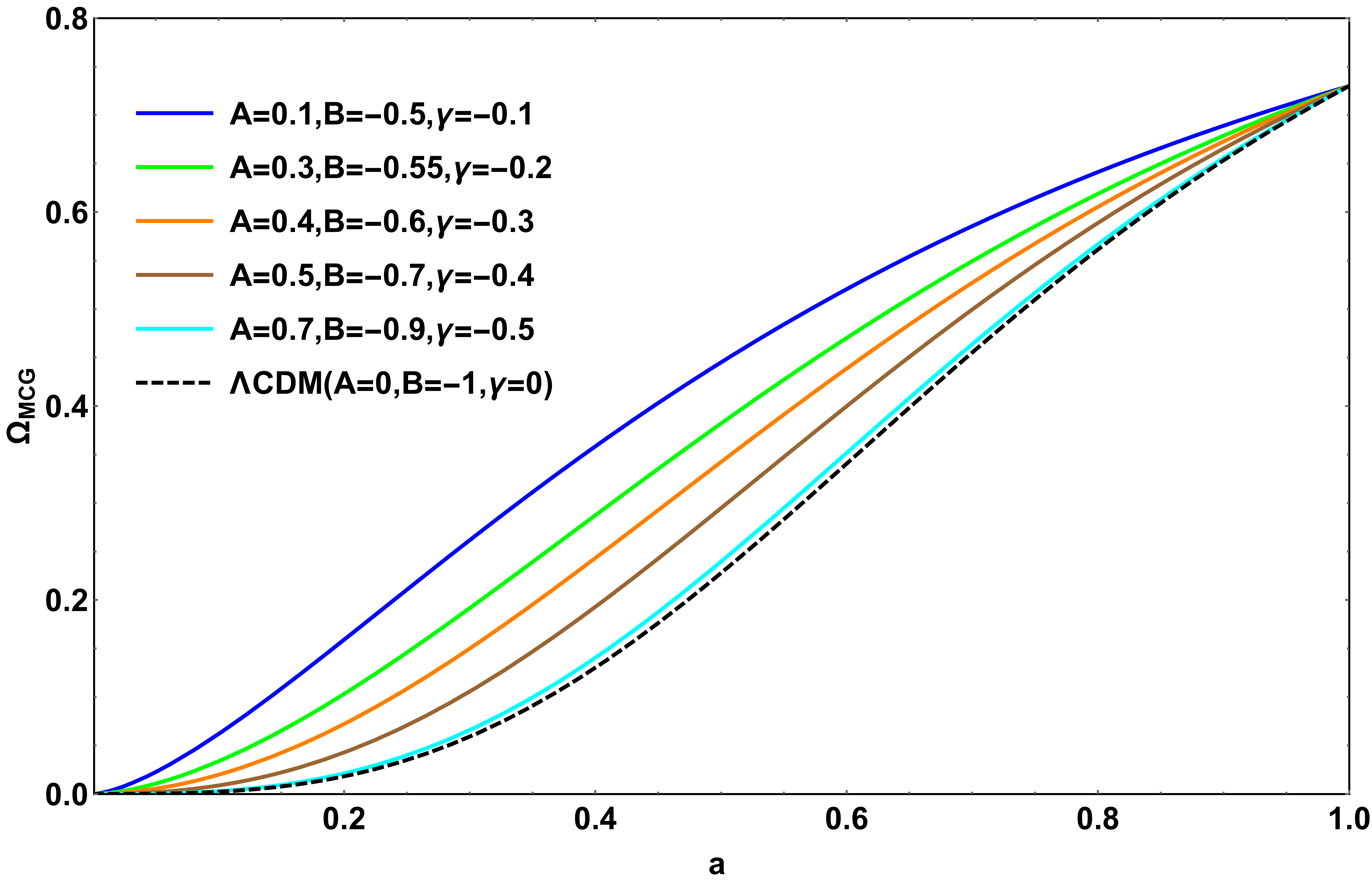}
  \includegraphics[width=7.5 cm]{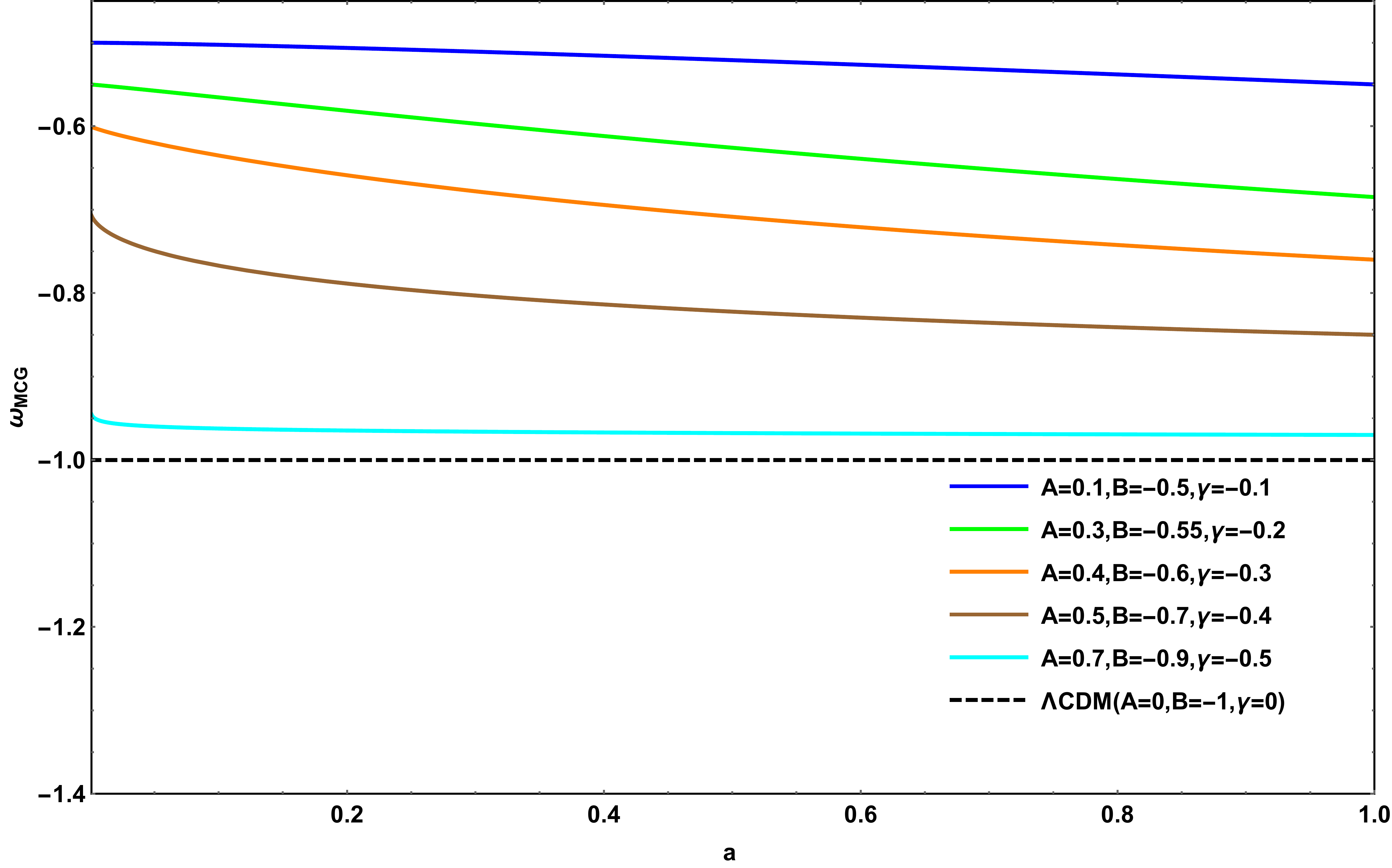}
  \includegraphics[width=7.5 cm]{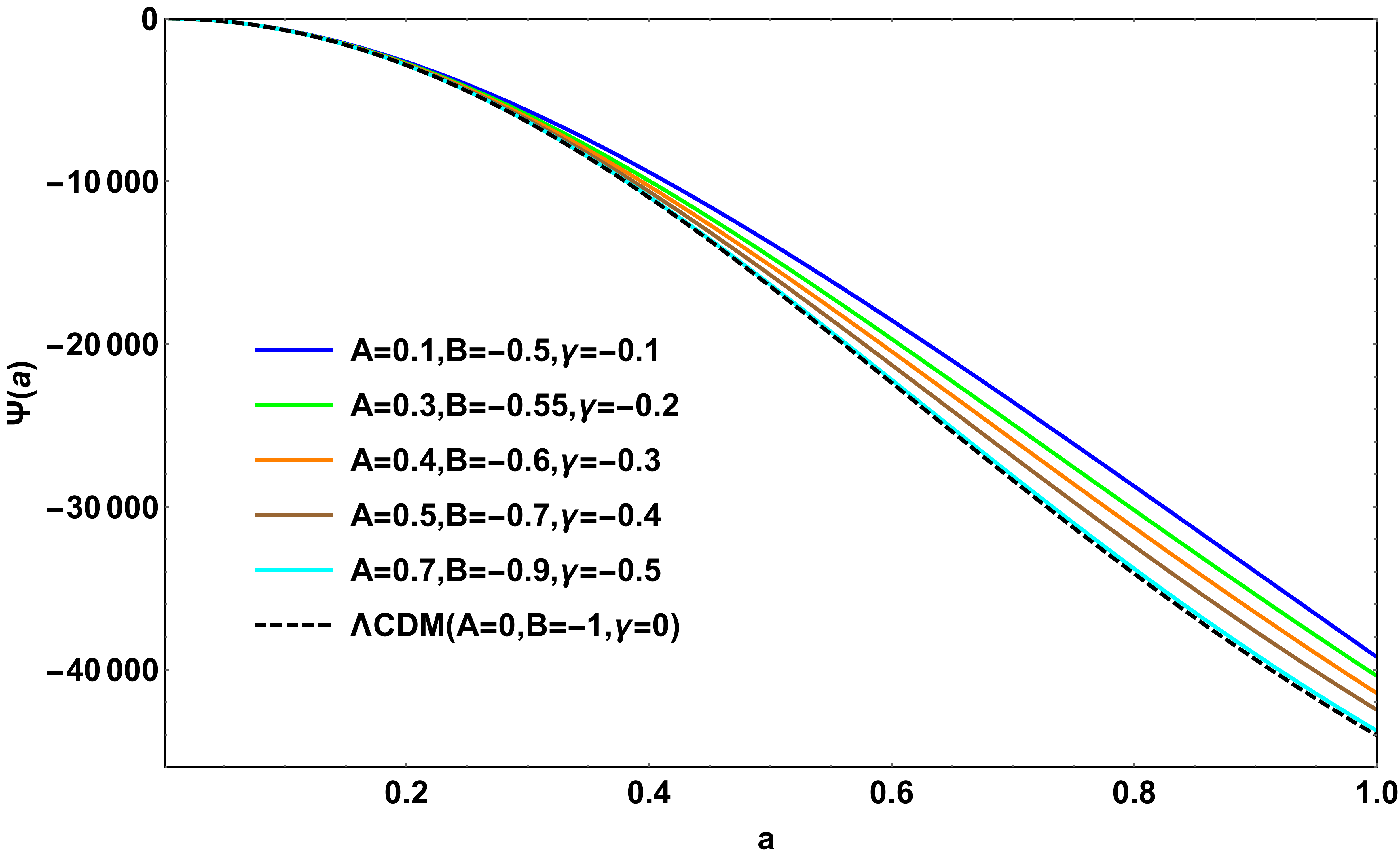}
  \includegraphics[width=7.5 cm]{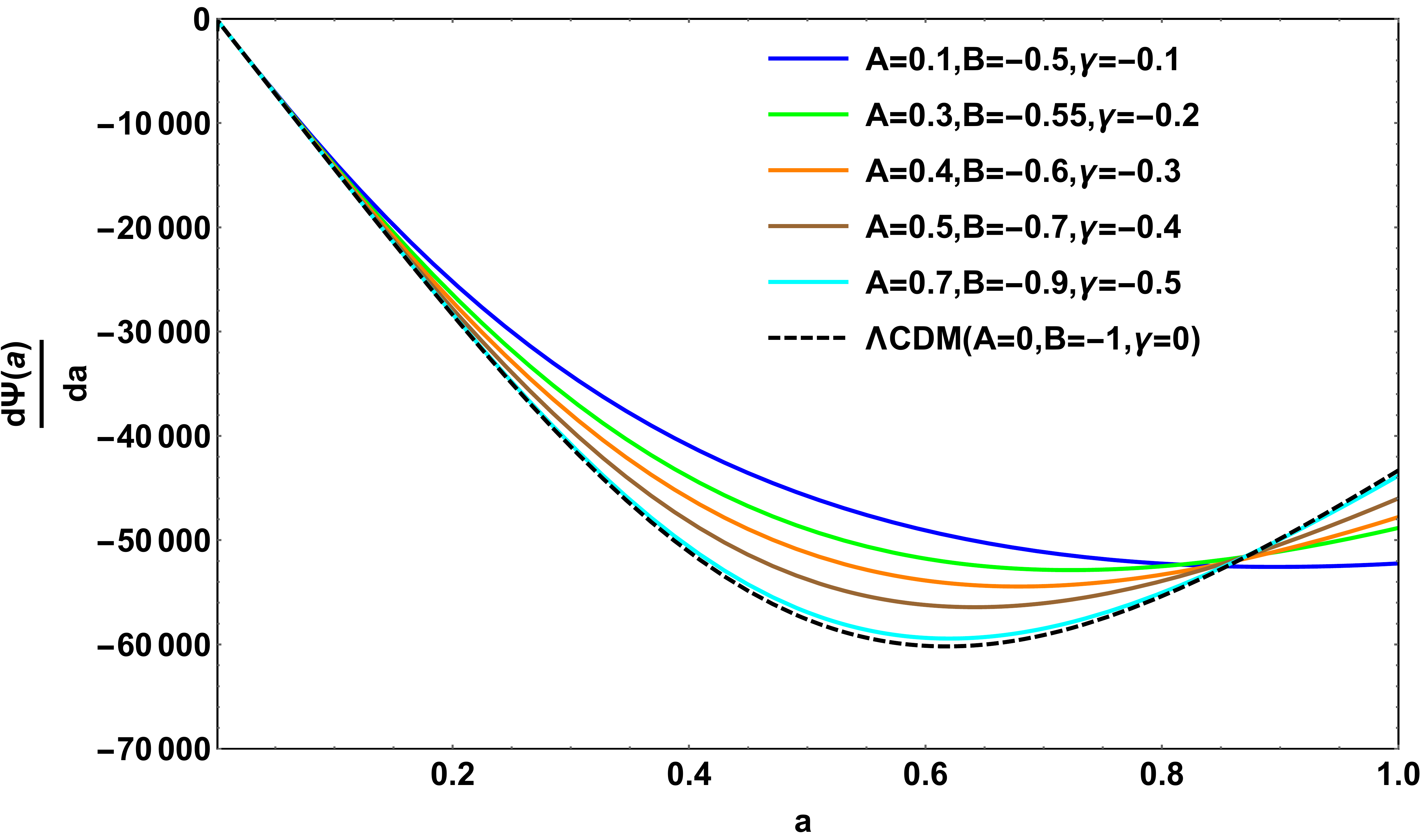}
\caption{Top left panel shows the  growth rate $f(a)$, top right panel shows the evolution of dark energy density parameter $\Omega_{MCG}$, middle left panel shows the evolution of the EoS parameter $\omega_{MCG}$, middle right panel shows the evolution of the configurational entropy $\Psi(a)$, and the bottom left panel shows the rate of change of configurational entropy ($\frac{d\Psi(a)}{da}$) for the MCG model.}
\label{FIG2}
\end{figure}

In Fig \ref{FIG1}, we show in the upper left panel, the evolution of growth rate $f(a)$ for the $\Lambda$CDM model and for the GCG model with various parameter combinations where we find that the suppression of growth rate is minimal for the $\Lambda$CDM model which correspond to $A=0$ and $B=-1$. For the GCG model, the suppression increases as we decrease the parameter $A$. Additionally, the parameter $B$ seem to behave oppositely wherein, a decrease in $B$ decreases the suppression of $f(a)$.  \\
In the upper right panel, we show the evolution of dark energy density parameter $\Omega_{GCG}$ where we find for the $\Lambda$CDM model, the profile is the steepest and this explains the minimal suppression of the growth rate $f(a)$. Additionally, as expected, a decrease $A$ or an increase in $B$ steepens the profiles.\\
The middle left panel shows the evolution of the EoS parameter $\omega_{GCG}$ where we find all the profiles to remain in the Quintessence region and start approaching the $\Lambda$CDM as $A$ increases or as $B$ decreases.\\
In the middle right panel, we show the evolution of the configurational entropy $\Psi(a)$ where we find the dissipation of $\Psi(a)$ to be the maximum for the $\Lambda$CDM model. Moreover, as the parameters $A$ decrease and $B$ increase, the dissipation lessens. The large dissipation of $\Psi(a)$ for the $\Lambda$CDM model can be attributed to the fact that $\Psi(a)$ is a derived quantity and hence depends chiefly on the evolution of the growth rate. Thence, since the growth rate of clustering is the maximum in the $\Lambda$CDM cosmology, it therefore harbor the tendency to dissipate the maximum $\Psi(a)$. \\ 
In the bottom left panel, we show the rate of change of configurational entropy ($\frac{d\Psi(a)}{da}$) where we find the minima transpire at a larger scale factor as $A$ decreases and $B$ increases. For the $\Lambda$CDM model, the minima occur at $a_{DE}\simeq 0.61$ which correspond to a redshift of transition of $z_{DE}\simeq 0.64$. For $\left\lbrace A,B\right\rbrace $ = $\left\lbrace 0.25,-0.4\right\rbrace$ , the minima hits at $a_{DE}\simeq0.89$ or at $z_{DE}=0.12$ which is not consistent with observations \cite{ratra}. Similarly, for the $\left\lbrace A,B\right\rbrace $ = $\left\lbrace 0.4,-0.5\right\rbrace $, the same transpire at $z_{DE}=0.4$ which is again observationally unfavorable. Nonetheless, for $\left\lbrace A,B\right\rbrace $ = $\left\lbrace 0.74,-0.69\right\rbrace $ and $\left\lbrace 0.8,-0.89\right\rbrace $, the minima occur at $z_{DE}=0.58$ and at $0.61$ respectively, both of which are consistent with observational constraints \cite{ratra}.\\
In Fig \ref{FIG2}, we show in the upper left panel, the evolution of growth rate $f(a)$ for the MCG model and compare it with the same obtained for the $\Lambda$CDM model (represented by $A=0$, $B=-1$, and $\gamma=0$) where we find that the profiles to lean towards the $\Lambda$CDM as the parameter $\gamma$ is lowered. Similar trends are also observed for the evolution of dark energy density parameter $\Omega_{MCG}$, dark energy EoS parameter $\omega_{MCG}$, and for the configurational entropy $\Psi(a)$. As the parameter $\gamma$ lessens, the rate of change of configurational entropy attains the minima at larger scale factors. For $\left\lbrace A,B,\gamma \right\rbrace $ = $\left\lbrace 0.1,-0.5,-0.1 \right\rbrace$, the minimum occur at $a_{DE}\simeq0.9$ which is not favorable. Nonetheless, for suitable parameter choices such as $\left\lbrace A,B,\gamma \right\rbrace $ = $\left\lbrace 0.5,-0.7,-0.4 \right\rbrace$ and $\left\lbrace A,B,\gamma \right\rbrace $ = $\left\lbrace 0.7,-0.9,-0.5 \right\rbrace$, we get $a_{DE}\simeq0.64$, and $0.62$ respectively and are in excellent agreement with observations. So for both the models, there subsist several parameter conjunctions that give rise to a healthy dissipation of configurational entropy and therefore produce minima which are observationally consistent with an accelerating Universe that started at a redshift $z_{DE}\simeq0.61$. 

\section{Conclusions}
Even though the $\Lambda$CDM cosmological model describes the dynamics of the Universe from the primordial times to the present accelerated phase elegantly, there exist several problems in cosmology which the model cannot explain. In this spirit, many alternate models have emerged and hitherto, Chaplygin gas models are one of the most promising candidates and have passed many observational tests. \\
In Ref \cite{19}, the rate of change of configurational entropy was proposed to achieve a minimum which largely depends on the cosmological model and indicate the epoch of an accelerated Universe. It may be noted that the relative dominance of the dark energy model under consideration in controlling the growth rate of large scale structures dictates the location of $a_{DE}$ \cite{20}. Upon careful inspection, one may find that the third term Eq. \ref{3} contain a distinct combination of the scale factor, growing mode, and its temporal derivative and it is largely because of such conjunctions of these entities that the first derivative of configurational entropy attains a minimum and represent the eon of dark energy domination and therefore acts a robust and straightforward method to impose constraints on a cosmological model.\\
In this work, we use the Linder parametrization of a constant growth index \cite{linder/index} to explore the evolution of growth rate of clustering and the dissipation of configurational entropy in some of the most widely studied Chaplygin gas models, such as the generalized Chaplygin gas and the modified Chaplygin gas. The model parameters for both the models play a decisive role in the evolution of growth rate, dark energy density parameter, EoS parameter, and configurational entropy. Additionally, the work also report that the rate of change of configurational entropy acquires a minimum which depend exclusively on the choice of model parameters and that there exists suitable parameter combinations giving rise to a viable dissipation of configurational entropy, and therefore confirming its time derivative to hit a minimum at a scale factor which complies with the current observational constraints on the redshift of transition from a dust to an accelerated Universe and thus making Chaplygin gas models a viable candidate for dark energy.

\section*{Acknowledgments}

We thank an anonymous reviewer for the helpful comments.

\end{document}